\documentstyle[twocolumn,epsf]{jpsj}
\def\runtitle{Magnetization Jump in the Spin-$\frac{1}{2}\ XXZ$ Chain with Next-Nearest-Neighbour Coupling}
\def\runauthor{Shunsaku {\sc Hirata}}
\def\lesssim{\mathrel{\mathpalette\vereq<}}
\def\vereq#1#2{\lower3pt\vbox{\baselineskip1.5pt \lineskip1.5pt
 \ialign{$#1\hfill##\hfil$\crcr#2\crcr\sim\crcr}}}
\title{%
Magnetization Jump in the Spin-$\mib{\frac{1}{2}}\ \mib{XXZ}$ Chain with
Next-Nearest-Neighbour Coupling
}
\author{%
Shunsaku {\sc Hirata}\footnote{E-mail: hira4scp@mbox.nc.kyushu-u.ac.jp}
}
\inst{Department of Physics, Kyushu University, Fukuoka 812-8581}
\recdate{\today}
\abst{%
The zero-temperature magnetization process of 
a one-dimensional $S=\frac{1}{2}$ spin system with next-nearest-neighbour
coupling $\alpha$ and anisotropy $\delta$ is studied.
There is a region in the parameter space where the magnetization
has a jump at certain external field.
The boundary of this region is determined.
The two cases are distinguished according as $\alpha<\frac{1}{4}$ or
$\alpha>\frac{1}{4}$.
In the latter case, the energy dispersion of the low-lying excitations
exhibits a double minimum structure.
}
\kword{quantum spin chain, magnetization jump, two magnon bound state,
frustration}
\begin{document}
\sloppy
\maketitle

The magnetization process is one of the interesting topics in 
the study of low-dimensional quantum spin systems.
In particular, the one-dimensional systems at low-temperature
attracted a great deal of theoretical and experimental interest
because of pronounced effect induced by quantum fluctuation
in those systems.
They exhibit various nontrivial behaviours, {\it e.g.},
the middle-field cusp singularity~\cite{OHA99},
the magnetization plateau~\cite{Hid94,Oka95} and
the second-order spin-flop transition.~\cite{ST99}
What we concern in the present study is the magnetization
jump similar to the metamagnetic transition in the classical
Ising-like anti\-ferro\-magnet, but it is caused by a different mechanism.

In this paper, we study a spin-$\frac{1}{2}\ XXZ$ chain with
nearest-neighbour (NN) coupling $J$ and next-nearest-neighbour (NNN)
coupling $J\alpha$ in a magnetic field $h$.
The model is described by the Hamiltonian
\begin{eqnarray}
H & = & H_0-h\sum_{i=1}^{L}S_i^z ,
\label{EQ.1}\\
H_0& = & J\sum_{i=1}^{L} (h_{i,i+1}+\alpha h_{i,i+2}),
\label{EQ.2}\\
h_{i,j} & = & S_i^xS_j^x+S_i^yS_j^y+\delta S_i^zS_j^z ,
\label{EQ.3}
\end{eqnarray}
where $\mib S_i$'s represent the $S=\frac{1}{2}$ spin operators on the
chain, and $L$ is the total number of spins.
The periodic boundary condition $\mib S_{i+L}=\mib S_i$ is assumed.
Hereafter, we set $J>0$ and take a unit so that $J=1$.
The model has been studied extensively. In the absence of external
fields, it exhibits a variety of phenomena associated with competing
interaction such as spontaneous dimerization~\cite{Hal82,ON92,NO94}
and incommensurate spin correlations.~\cite{TH87,WA96} 

\begin{figure}[tb]
\leavevmode\centering
\epsffile{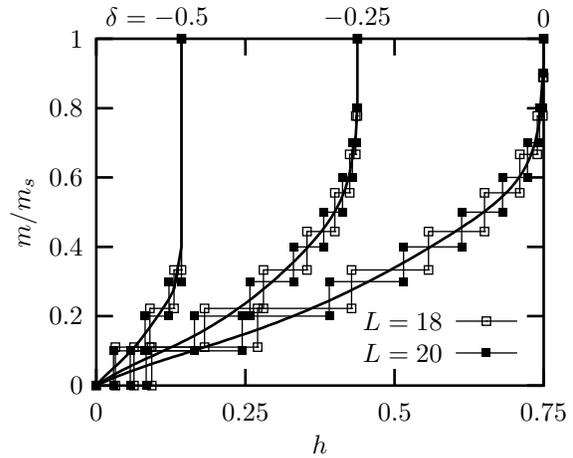}
\caption{The zero-temperature magnetization curves
for $\alpha=0.25$ and $\delta=0.0,-0.25,-0.5$.
The thin lines show the finite-size results for $L=18$ and $20$.
The magnetization (per site) $m$ is normalized by
the saturation magnetization $m_s=\frac{1}{2}$.
The thick lines represent the estimated limiting ($L\to\infty$)
curves. For $\delta=-0.25$ and $-0.5$, there exists a critical field
$h_c$ at which the magnetization jumps to the saturation value $m_s$.}%
\label{FIG.1}
\end{figure}
We shall concentrate on the region
$0\le\alpha\le\frac{1}{2},-1<\delta<0$ where an anomalous
behaviour in the magnetization process was reported~\cite{GMK98}:
In a region of parameter space, namely,
$\delta_f(\alpha)<\delta<\delta_c(\alpha)$,
there exists a critical field $h_c$
at which the magnetization jumps to its maximum value $m_s=\frac{1}{2}$
(see Fig.~\ref{FIG.1}).
It should be noted that the model has competing NN and NNN exchange
couplings in the easy-plane ($XY$-plane), while there is no competition in
the $z$-axis although couplings are relatively weak.
It is responsible for the existence of the magnetization jump:
The system prefers to oder in the $XY$-plane but NNN coupling introduces
frustration.
Therefore, it becomes energetically preferable to order along the $z$-axis
for sufficiently large $\alpha$ and $|\delta|$.
For $\delta<\delta_f$, the zero-field ground state becomes
the completely aligned (ferro\-magnetic) state with $m=\frac{1}{2}$.

The purpose of this paper is to determine the boundary
$\delta=\delta_c(\alpha)$ of the region where the magnetization
jump takes place.~\cite{NOTE1}

\begin{figure}[t]
\leavevmode\centering
\epsffile{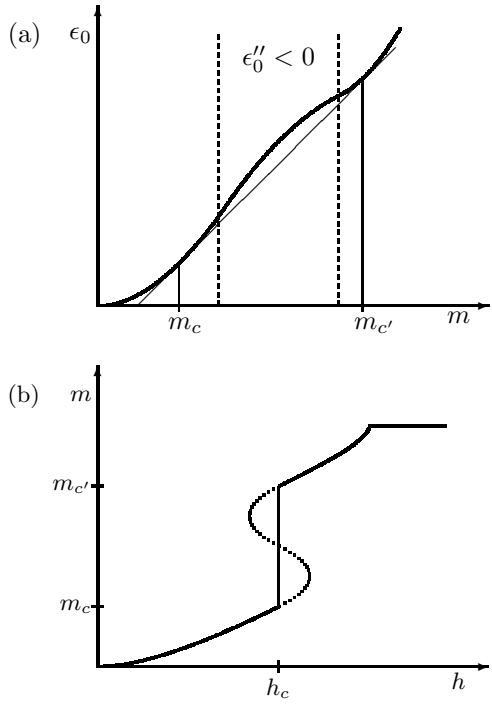}
\caption{(a) Schematic curve of $\epsilon_0(m)$ in the case when the
magnetization jump occurs. There is a region in which $\epsilon_0''$ is
negative. The line with slope $h_c$ is also shown where $h_c$ is a
critical field. (b) Magnetization curve corresponding to $\epsilon_0$ of (a).
}%
\label{FIG.2}
\end{figure}

First, we discuss the condition that the magnetization jump occurs.
For the magnetization curve to be smooth,
$\epsilon_0(m)$, the lowest energy eigenvalue (per site) for
the Hamiltonian $H_0$ in
the fixed $m=\sum S^z_i/L$ sector, must be monotonically increasing in $m$
and convex for $0< m< m_s=\frac{1}{2}$;
\begin{equation}
\epsilon_0''(m)=\frac{d^2\epsilon_0(m)}{dm^2}>0\ .
\label{EQ.4}
\end{equation}
On the other hand, if there exists a region where 
$\epsilon_0''\le 0$, then the magnetization jumps over that region as
schematically shown in Figs.~\ref{FIG.2}(a) and~\ref{FIG.2}(b).
Thus, the presence of such a region as $\epsilon_0''\le 0$
means that the magnetization jump occurs.
The quantity $\epsilon_0''(m)$ can be estimated from 
\begin{equation}
\epsilon_0''(m)=\lim_{L\to\infty}L\times \Delta(m) ,
\label{EQ.5}
\end{equation}
where
\[
\Delta(m)=E\left(m+\frac{1}{L}\right)
+E\left(m-\frac{1}{L}\right)-2E(m) .
\]

\begin{figure}[t]
\leavevmode\centering
\epsffile{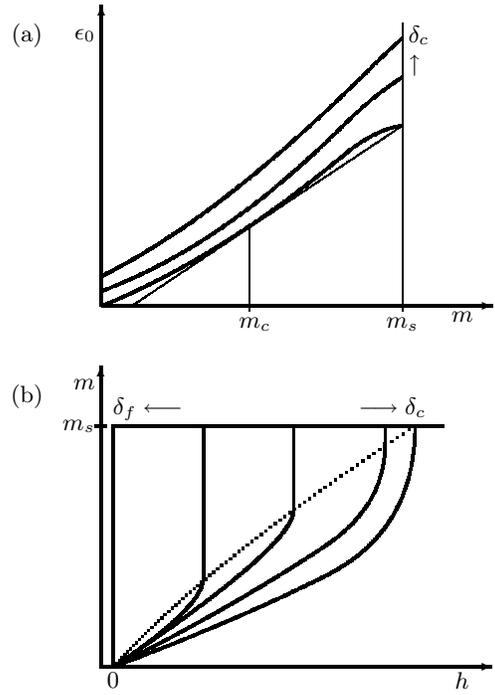}
\caption{(a) The $m$-dependence of $\epsilon_0$ and
(b) magnetization curves of the $S=\frac{1}{2}\ XXZ$ chain with
NNN interaction in the case when $\delta_f\le\delta\le\delta_c$.
The dotted line indicates the critical magnetization $m_c(\delta)$.
}%
\label{FIG.3}
\end{figure}
In the case of our model, the problem of
determining the boundary curve $\delta=\delta_c(\alpha)$
is simplified.
The results of the exact diagonalization suggest that
the critical magnetization $m_c$ increases as $\delta$ is increased
and reaches $m_s$ at $\delta=\delta_c$.
It indicates that
\begin{equation}
\epsilon_0''|_{m=\frac{1}{2}-0}\le 0
\label{EQ.6}
\end{equation}
at least for $\delta\lesssim\delta_c$
[see Figs.~\ref{FIG.1},~\ref{FIG.3}(a) and \ref{FIG.3}(b)].
The boundary $\delta_c$ is
a line where $\epsilon_0''|_{m=\frac{1}{2}-0}$ changes sign.
Therefore, we have merely to consider up to two-spin deviations from the
completely aligned state for our purpose.
We put $\Delta^{(1)}=\Delta(\frac{1}{2}-\frac{1}{L})$;
\begin{equation}
\Delta^{(1)}=E_0^{(2)}-E_0^{(0)}
-2\left[E_0^{(1)}-E_0^{(0)}\right] ,
\label{EQ.7}
\end{equation}
where $E_0^{(0)}$ denotes the energy of the ferro\-magnetic
state and $E_0^{(n)}$ the lowest energy of the states with $n$-spin
deviations from the ferro\-magnetic state.
For $\delta>\delta_f$, the ferro\-magnetic state is not a ground state
of the Hamiltonian $H_0$. However, we shall call those $n$-deviation
states as $n$-magnon states
although they are not the excitations from the ground state.

\begin{figure}[tb]
\leavevmode\centering
\epsffile{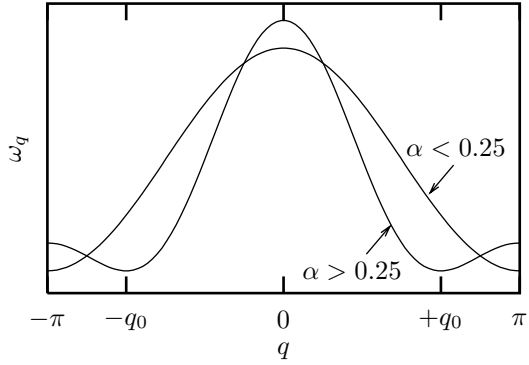}
\caption{Energy spectrum of one-magnon states. For $\alpha>\frac{1}{4}$,
two minima appears at $q=\pm q_0$
where $q_0=\arccos(-\frac{1}{4\alpha})$.}
\label{FIG.4}
\end{figure}

We now obtain the energy spectrum of one- and two-magnon states.
We rewrite the Hamiltonian (\ref{EQ.2}) in term of fermion
creation and annihilation operators with momentum $k$, $a_k^+$ and $a_k$,
via Jordan-Wigner transformation. The Hamiltonian becomes
\begin{eqnarray}
H&=&E_0^{(0)}+\sum_{k}\omega_qa_q^+a_q\nonumber\\
&&-\frac{1}{L}\sum_{1,2,3,4}\delta_{1+2,3+4}\Gamma_{1,2,3,4}a_1^+a_2^+a_3a_4 ,
\label{EQ.8}
\end{eqnarray}
where
\[
\left \{ \begin{array}{ccl}
E_0^{(0)}&=&\frac{1}{4}(1+\alpha)\delta L\\
\omega_q&=&\cos q-\delta+\alpha(\cos 2q-\delta) \\
\Gamma_{1,2,3,4}&=&\delta\cos(q_2-q_4)+\alpha\delta\cos\left(2(q_2-q_4)\right)\\
&&+2\alpha\cos (2q_1+q_2-q_3)\\
\end{array}
\right.,
\]
and we have used shorthand notations $1,2,3,4$ for
$q_1,q_2,q_3,q_4$.
In eq.~(\ref{EQ.8}), $E_0^{(0)}$ and $\omega_q$ represent
the energy of the ferro\-magnetic state $|0\rangle$
and the excitation spectrum of the one-magnon state
with momentum $q$, respectively.
For $\alpha\le\frac{1}{4}$, $\omega_q$ has a single minimum at $q=\pi$,
while, for $\alpha >\frac{1}{4}$,
the minimum shifts from $q=\pi$ and two minima appears at $q=\pm q_0$
where $q_0=\arccos(-\frac{1}{4\alpha})$ (Fig.~\ref{FIG.4}).

The total wave number $Q$ of two magnons to be conserved, thus we write
two-magnon state as
\begin{equation}
|Q\rangle =\sum_{q>0} f_Q(q) a_{q_1}^+a_{q_2}^+ |0\rangle ,
\label{EQ.9}
\end{equation}
where
\[
Q=q_1+q_2 \ , \ q =\frac{1}{2}(q_1-q_2)\ . 
\]
Let us solve the eigenvalue equation
\begin{equation}
H|Q\rangle=E|Q\rangle ,
\label{EQ.10}
\end{equation}
where $E$ is the energy eigenvalue of the two-magnon state.
Taking the inner product of
both side of eq.~(\ref{EQ.10})
with every basis state $\langle 0|a_{q_1}a_{q_2}$,
we obtain the following equation:
\begin{eqnarray}
[E-\gamma_{Q,q}] f_Q(q)=\frac{4}{L}\sum_{\kappa>0}f_Q(\kappa)
[\chi_Q\sin q\sin\kappa\nonumber\\
+\alpha\delta \sin 2q\sin 2\kappa] ,
\label{EQ.11}
\end{eqnarray}
where
\[
\left\{\begin{array}{ccl}
\gamma_{Q,q} &=& E_0^{(0)}+\omega_{q_1}+\omega_{q_2} \\
\chi_Q &=&\delta +2\alpha\cos Q\\
\end{array}\right.\ . 
\]
Here, $\gamma_{Q,q}$ represents the energy of the free two-magnons
with total wave number $Q=q_1+q_2$, and forms
a continuum for each value of $Q$.

We are interested in whether there exists a solution of eq.~(\ref{EQ.11})
which gives $\Delta^{(1)}<0$.
Hence, it is sufficient to look for the solution satisfies
$E<\gamma_{Q,q}$, i.e., the two-magnon bound states below the 
continuum [cf. eq.~(\ref{EQ.7})].
Then, the eigenvalue equation reduces to
(see, for instance, ref.~\citen{Ono71})
\begin{equation}
D_Q(E)=
\left|\begin{array}{cc}
1-2\chi_Q I^{2,0}_Q & -2\alpha\delta I^{1,1}_Q\\
\\
-2\chi_Q I^{1,1}_Q & 1-2\alpha\delta I^{0,2}_Q
\end{array}
\right| =0 \ ,
\label{EQ.12}
\end{equation}
where
\begin{eqnarray}
I^{l,m}_Q(E)&=&\frac{2}{L}\sum_{q>0}\frac{\sin^lq\sin^m2q}{E-\gamma_{Q,q}} \nonumber \\
&\stackrel{L\to\infty}{\longrightarrow}&
\frac{1}{\pi}\int_0^{\pi}dq\frac{\sin^lq\sin^m2q}{E-\gamma_{Q,q}} .
\label{EQ.13}
\end{eqnarray}
It is the sufficient and necessary condition for the bound state
lying outside of the continuum to exist; i.e.,
such bound states exists if and only if eq.~(\ref{EQ.12}) folds.
The explicit solution of (\ref{EQ.12}) is found analytically
for several values of $\alpha,\delta$ and $Q$.
It is investigated numerically for general case.

\begin{figure}[tb]
\leavevmode\centering
\epsffile{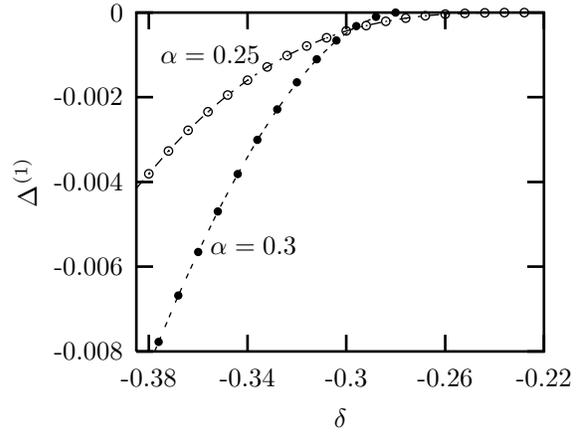}
\caption{The results of $\Delta^{(1)}$ for $\alpha=0.25$ and $\alpha=0.3$.
There is a point $\delta_c$ where
$\Delta^{(1)}$ vanishes, which is a boundary of the region where
the magnetization jump occurs.
}
\label{FIG.5}
\end{figure}
\begin{figure}[tb]
\leavevmode\centering
\epsffile{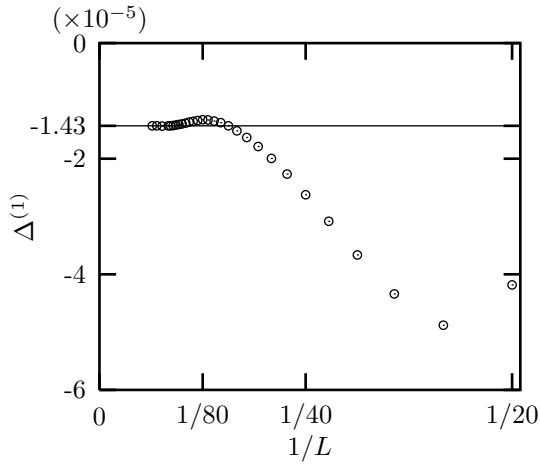}
\caption{Comparison with the result from the numerical diagonalization.
Open circles denotes the finite chain results for $L=20\mbox{-}156$.
The result obtained from eq.~(\ref{EQ.12}) is
$\Delta^{(1)}=-1.43\times 10^{-5}$.}
\label{FIG.6}
\end{figure}

Figure \ref{FIG.5} shows the results of $\Delta^{(1)}$
for $\alpha=0.25$ and $0.3$.
It can be seen that there exists a point $\delta_c$
at which $\Delta^{(1)}$ vanishes for each value of $\alpha$
and that $\Delta^{(1)}$ is negative for $\delta<\delta_c$ which suggests
the magnetization jump to occur.~\cite{NOTE3}
In Fig.~\ref{FIG.6}, the result is compared with the one obtained from
the diagonalization of the finite chain Hamiltonian for rather
large systems ($L=20\mbox{-}156$) at $\alpha=0.25,\delta=-0.25$.
Note that there is a weak oscillatory behaviour in the finite-size
results.
For $\alpha>\frac{1}{4}$, there appears rapid oscillation,
thus it becomes difficult to estimate a limiting value of
$\Delta^{(1)}$ from the finite chain calculation.

\begin{figure}[t]
\leavevmode\centering
\epsffile{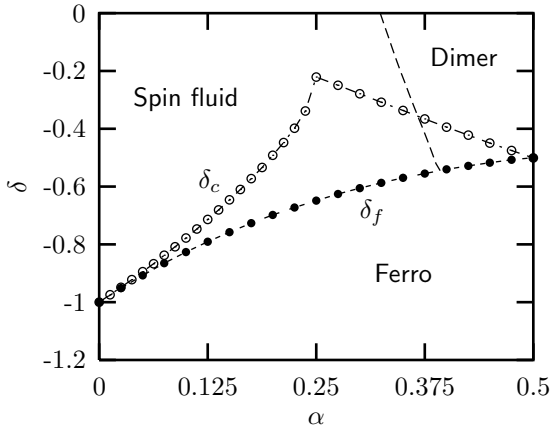}
\caption{The phase diagram in the $\alpha\mbox{-}\delta$ plane.
The magnetization jump occurs in the region enclosed by the two curves
$\delta_c$ ($\circ$) and $\delta_f$ ($\bullet$).
The zero-field ground-state phase diagram is also shown;
the boundary of the ferro\-magnetic phase $\delta_f$
and the dimer-spin fluid phase boundary (dashed line).
}
\label{FIG.7}
\end{figure}
In Fig.~\ref{FIG.7}, the curve $\delta_c(\alpha)$ is shown in the
$\alpha\mbox{-}\delta$ plane.
The magnetization jump occurs in the region enclosed by the two curves
$\delta_c$ and $\delta_f$.~\cite{NOTE4} 
The ground-state phases in the absence of
external fields are also shown;~\cite{HN99} there are three phases,
namely, the spin fluid phase which is characterized by the gapless excitation
and critical behaviour, the spontaneously dimerized phase
and the ferro\-magnetic phase.
From Fig.~\ref{FIG.7}, it can be clearly seen that
the curve $\delta_c$ has a kink at $\alpha=\frac{1}{4}$.
It results from the fact that the dispersion curve of
the one-magnon states changes its shape at $\alpha=\frac{1}{4}$:
For $\alpha<\frac{1}{4}$, it has a single minimum
at $q=\pi$, while, for $\alpha>\frac{1}{4}$, two minima appears.
The lowest lying state should consist of bound pairs of magnons
with $q\simeq\pi$ for the former case
and with $\pm q_0$ for the latter case.

To summarize, we determined the boundary of the region
in which the magnetization jump occurs in the $S=\frac{1}{2}\ XXZ$
chain with easy-plane anisotropy and NNN interaction
in the case of $0<\alpha<\frac{1}{2}$.
It was shown that the $\alpha$-dependence of the boundary curve $\delta_c$
is qualitatively different according as $\alpha<\frac{1}{4}$
or $\alpha>\frac{1}{4}$.
For $\alpha>\frac{1}{4}$, the region that the magnetization jump occurs
shrinks with increasing $\alpha$
in contrast to the case of $\alpha<\frac{1}{4}$.
It reflects the fact that 
the energy dispersion curve of the low-lying excitation
develops the double minimum structure beyond the line $\alpha=\frac{1}{4}$.

One might be interested in how the system is in the case
when the external field $h$ is close to the critical field $h_c$.
One of the possible situation is that
there are clusters of up-spins separated by fluid-like domains.

%%% ACKNOWLEDGMENT %%%
The author would like to thank K.~Nomura for valuable discussion and
comments.

%%% REFERENCES %%%


\begin{thebibliography}{99}
\bibitem{NOTE2}
The integrand of (\ref{EQ.13}) has, however,
a sharp peak which becomes singular as $E$ approaches to
the lower edge of the scattering continuum $\gamma_{Q,q}$.
Thus, for numerical quadrature we use
the double exponential formula,~\cite{TM74}
which is one of the most optimal and suitable numerical quadrature
formula for improper integrals.
\bibitem{NOTE3}
To discuss thermodynamic limit,
we must take into account the finite size correction:
If $\Delta^{(1)}$ behaves as
$\Delta^{(1)}=\Delta^{(1)}_{\infty}+c_1L^{-1}+O(L^{-2})$
and if $\Delta^{(1)}_{\infty}=0$,
then it is the coefficient of the correction term $c_1$
that determine the sign of $\epsilon_0''$ in thermodynamic limit.
We are unable to estimate the finite size correction analytically.
However, it seems that there is no significant difference in $\delta_c$.
\bibitem{NOTE4}
The boundary of the ferro\-magnetic phase $\delta_f$ was determined from
the level crossing of the ferro\-magnetic state and
the lowest eigenstate with $m=0$ (see ref.~\citen{HN99} for details).
\end{thebibliography}
\end{document}